\documentclass[aps,prb,twocolumn,superscriptaddress]{revtex4}

\usepackage{graphicx}
\usepackage{amssymb}
\usepackage{bm}

\begin{document}



\title{Electron-hole contribution to the apparent $s$-$d$ exchange interaction
in III--V diluted magnetic semiconductors}



\author{Cezary \'Sliwa}
\email{sliwa@ifpan.edu.pl}
\affiliation{Institute of Physics, Polish Academy of Sciences, al.\
  Lotnik\'ow 32/46, PL 02-668 Warszawa, Poland}

\author{Tomasz Dietl}
\email{dietl@ifpan.edu.pl}
\affiliation{Institute of Physics, Polish Academy of Sciences, al.\
  Lotnik\'ow 32/46, PL 02-668 Warszawa, Poland}
\affiliation{Institute of Theoretical Physics, University of Warsaw,
ul.\ Ho\.za 69, PL 00-681 Warszawa, Poland and
ERATO Semiconductor Spintronics Project,
  Japan Science and Technology, al.\ Lotnik\'ow 32/46,
  PL 02-668 Warszawa, Poland}


\date{27 November 2008}

\begin{abstract}
Spin splitting of photoelectrons in $p$-type and electrons in $n$-type III--V Mn-based diluted magnetic semiconductors is studied theoretically.
It is demonstrated that the unusual sign and
magnitude of  the apparent $s$-$d$ exchange integral
reported for GaAs:Mn arises from
exchange interactions between electrons and holes bound to Mn acceptors.
This interaction dominates over the coupling between electrons
and Mn spins, so far regarded as the main source of spin-dependent phenomena.
A reduced magnitude of the apparent $s$-$d$ exchange integral found in $n$-type materials is explained by the presence of repulsive Coulomb potentials at ionized Mn acceptors and a bottleneck effect.
\end{abstract}


\maketitle

\section{Introduction}

Owing to the possibility of gradual incorporation of
magnetism to the well-known semiconductor matrices, diluted magnetic
semiconductors (DMSs) (Refs. \onlinecite{Furdyna:1988}, \onlinecite{Dietl:1994},
\onlinecite{Matsukura:2002}, \onlinecite{Jungwirth:2006}) offer unprecedented opportunity for
examining {\em quantitatively} the origin of spin dependent couplings
between band carriers and electrons localized on the open $d$-shell.
According to thorough studies of Mn-based II--VI DMSs, the spin-dependent coupling of the band-edge electrons and
Mn spins is characterized by $N_0\alpha  =250 \pm 60$~meV,\cite{Dietl:1994} where
$N_0$ is the cation concentration and $\alpha$ is the $s$-$d$ exchange
integral. The above value of $N_0 \alpha$ is in a full accord with the notion that
spin-dependent effects in the conduction band of a tetrahedrally coordinated DMS originate from
the intra-atomic potential $s$-$d$ exchange interaction. Indeed,  the corresponding $s$-$d$
exchange energy is 392~meV in the case of free Mn$^{1+}$ ions,\cite{Dietl:1994_a} and in
a DMS it is a subject of up to twofold reduction by a covalent admixture of the anion $s$-type wave function to the
Kohn-Luttinger amplitude at the conduction-band edge. In accord to this insight, $N_0 \alpha = 0.3$~eV results from {\em ab initio} computations for $n$-(Ga,Mn)As.\cite{Dalpian:2006}

Surprisingly, the recent comprehensive studies of quantum wells of highly dilute paramagnetic ${\rm Ga}_{1-x} {\rm Mn}_x {\rm As}$
($x \leq 0.13 \%$) suggest {\em antiferromagnetic}
$N_0\alpha   = -23 \pm 8 \, \rm meV$ (Refs.~\onlinecite{Myers:2005} and~\onlinecite{Poggio:2005})
or $ -20 \pm 6 \, \rm meV$ (Ref.~\onlinecite{Stern:2007})
for photoelectrons at the band edge.
These observations have not been explained by the recent theory,\cite{Dalpian:2006}
and appear to challenge the time-honored notion that the spin-dependent
coupling between the electrons and Mn spins in a tetrahedrally coordinated DMS originates from
the necessarily ferromagnetic intra-atomic potential $s$-$d$ exchange.

The starting point of our approach is the realization that
the density of Mn acceptors in the studied\cite{Poggio:2005,Stern:2007} quantum wells of GaAs
was more than one order of magnitude lower
than the critical value corresponding to the insulator-to-metal
transition and the onset of the hole-mediated ferromagnetism in this system.
Furthermore, a relatively high growth temperature resulted
in a small concentration of compensating defects. Accordingly, the
conduction-band photoelectrons interacted with complexes consisting of both Mn
and hole spin, $d^5 + h$, which are bound by the electrostatic potential and mutually coupled
by a strong antiferromagnetic $p$-$d$ exchange interaction. We develop here theory of the exchange interaction for such a case
and show that it explains, with no adjustable parameters, the sign reversal
of the apparent $s$-$d$ exchange integral.
Furthermore, we demonstrate
that an assumption about the heating of the Mn spin subsystem, invoked in order to
describe the observed dependence of electron spin splitting on the
magnetic field,\cite{Myers:2005,Poggio:2005}
can be relaxed within the present theory.

Independently, much reduced spin splitting has been
found for electrons injected to InAs quantum dots containing
a neutral Mn acceptor.\cite{Kudelski:2007} This observation is consistent
with the invoked here mutual cancelation of the
$s$-$d$ and $s$-$p$ exchange energies.

Another case where the presence of bound holes is of primary importance is the Bir-Aronov-Pikus relaxation of electron spins.
Surprisingly, it has recently been found\cite{Astakhov:2008} that the electron spin relaxation time in GaAs:Mn is by two orders of magnitude longer comparing to GaAs:Ge, challenging a general belief that magnetic impurities are efficient spin coherence killers. This puzzling observation
has been successfully interpreted\cite{Astakhov:2008} in accord to the theory presented here.\cite{Sliwa:2007}

While our model elucidates the origin of the anomalous sign and magnitude of the apparent $s$-$d$ exchange integral for photoelectrons in $p$-type DMSs, it does not explain a reduced magnitude of this energy observed by electron-spin resonance in GaN:Mn,\cite{Wolos:2003} and by electron spin-flip Raman scattering in GaAs:Mn.\cite{Heimbrodt:2001} We examine also this issue and demonstrate that the presence
of a bottleneck effect and of a repulsive potential associated with ionized Mn acceptors in compensated III--V Mn-based DMSs leads to a sizable, Mn concentration dependent, reduction of the $s$-$d$ exchange integral.

Our paper is organized as follows.
In Sec.~II we discuss a comparison of our theoretical results to experimental findings, delegating a detail description of the theory to subsequent sections.
Thus, in Sec.~III we present the adopted model of the Mn acceptor in GaAs, including the form of the envelop functions and relevant Land\'e factors. This is followed by the derivation of the exchange integral $J_{\mathrm{eh}}$ describing the spin-dependent interaction between band electrons and bound holes, considering first the short-range (Sec.~IV)
and then the long-range part (Sec.~V) of the electron-hole coupling.
Finally, in Sec.~VI we examine the effect of compensation on the magnitude of the apparent $s$-$d$ exchange integral. Section VII contains a summary and outlook.

An important aspect of our theory is that the exchange integrals describing the coupling between conduction-band electrons and holes bound to acceptors can be expressed, with no adjustable parameters, by the acceptor envelop functions $f(r)$ and $g(r)$ as well as by
the exchange splitting $\Delta$ and the longitudinal-transverse splitting $\Delta_{LT}$ of the bulk free excitons.

\section{Explanation of the observations}

The Mn acceptor complex can be described within the tight-binding approximation\cite{Yakunin:2004} or in terms of the Baldareschi-Lipari spherical model as proposed
by Bhattacharjee and Benoit \`a la Guillaume\cite{Bhattacharjee:2000} for GaAs:Mn and more recently employed to study impurity band effects.\cite{Fiete:2005}
We determine within this model how polarizations of a Mn spin $S =5/2$ and
of a hole total angular momentum $J=3/2$ depend on the magnetic field $B$ and temperature  $T$.
We then derive the form and magnitude of the exchange interactions between conduction-band
electrons and holes bound by Mn acceptors, extracting relevant electron-hole $s$-$p$ exchange
parameters from the previous experimental studies of the free exciton in GaAs.

The Mn acceptor Hamiltonian for the magnetic field along $z$ direction reads,
\begin{equation}
  \mathcal{H} = \varepsilon \bm{J} \cdot \bm{S} + \mu_B B (g_{\rm Mn} S_z + g_h J_z),
\end{equation}
where
$\varepsilon = 5 \, \rm meV$ is the experimentally determined
$p$-$d$ exchange energy in the Mn acceptor,\cite{Bhattacharjee:2000}
$g_{\rm Mn} = 2.0$, and $g_h = 0.75$ the is hole Land\'e factor
derived in Sec.~III.  From
the corresponding density matrix $\varrho = \exp[-\mathcal{H}/(k_B T)]$
we obtain $\left<S_z\right>_{T,B}$ and $\left<J_z\right>_{T,B}$.
Within the molecular-field approximation, the exchange splitting of the conduction-band edge for uncompensated GaAs:Mn becomes
\begin{equation}
\hbar \omega_s(T,B) = x N_0 [-\alpha \left<S_z\right>_{T,B} + J_{\mathrm{eh}}\left<J_z\right>_{T,B}],
\end{equation}
where the second term arises from the coupling $(J_{\mathrm{eh}}/\mathcal{V}) \, \bm{s} \cdot \bm{J}$
between the spin $\bm{s}$ of a band-edge electron and the
angular momentum $\bm{J}$ of holes bound to Mn acceptors ($\mathcal{V}$ is the volume of the sample). As shown in Secs.~IV and V, this interaction is
characterized by the $s$-$p$ exchange energy $N_0J_{\mathrm{eh}} = -0.51 \pm 0.17 \, \rm eV$. Hence, for the expected values of $N_0\alpha$, the $s$-$p$ exchange dominates over the $s$-$d$ interaction. Furthermore,  because of an antiferromagnetic sign of the $p$-$d$ exchange interaction, $\left<S_z\right>_{T,B}/\left<J_z\right>_{T,B} < 0$,
the apparent coupling between the electron and Mn complex is antiferromagnetic. In particular, adopting $N_0\alpha  = 0.219$~eV we obtain the field dependence of electron spin splitting shown in Fig.~1.

\begin{figure}[bt]
  \centering
  \includegraphics[width=0.95\columnwidth]{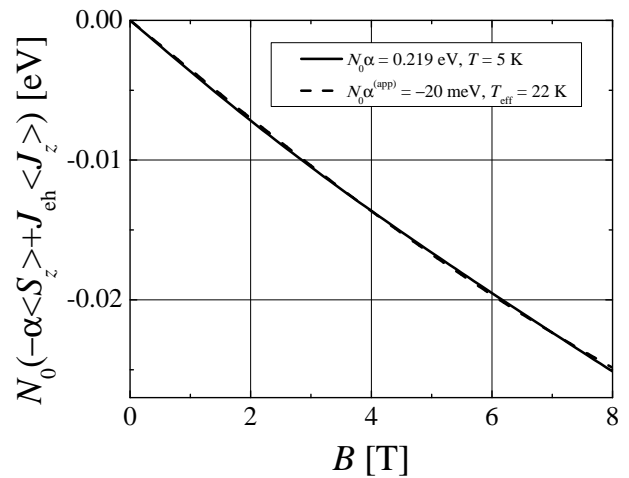}
  \caption{Theoretical values of the electron spin-splitting energies $\hbar \omega_s(B)/x$ (solid line)
  computed as a function of the magnetic field at 5~K.
  Dashed line represents fitting to the solid line obtained
  by treating the apparent $s$-$d$ exchange energy $N_0 \alpha^{\mathrm{(app)}}$ and temperature $T_{\mathrm{eff}}$ as adjustable parameters
  within the model that neglects the
  presence of the electron-hole exchange interaction $(J_{\mathrm{eh}}=0)$.}
  \label{fig1_eh}
\end{figure}

We recall that  the data on the photoelectron precession
frequency\cite{Myers:2005,Poggio:2005,Stern:2007} were interpreted
neglecting the presence of the bound holes $(J_{\mathrm{eh}}=0)$ as well as by treating both $N_0\alpha$ and $T$ in the Brillouin
function $\mathrm{B}_S(T,B)$ for $S = 5/2$ as adjustable parameters.\cite{Myers:2005,Poggio:2005,Stern:2007}
Proceeding in the same way we can describe our theoretical results very well with $N_0 \alpha^{\mathrm{(app)}} = -20\, \rm meV$,
$T_{\mathrm{eff}} = 22\, \rm K$, as shown by dashed line in Fig.~1.
We see that the present theory  explains why the small antiferromagnetic
apparent exchange energy $N_0 \alpha^{\mathrm{(app)}} = -20\pm 6\,
\rm meV$ and enhanced temperature $T_{\mathrm{eff}} = 20 \pm 10\, \rm K$ were found
experimentally.\cite{Poggio:2005,Stern:2007}

It is worth noting that if the contributions of the two terms determining spin splitting compensate each other, the fitted values of  $N_0 \alpha^{\mathrm{(app)}}$ and $T_{\mathrm{eff}}$
become correlated, so that only their ratio can be determined accurately.  However, this correlation affects little the experimentally determined band-edge value of $N_0\alpha^{\mathrm{(app)}} $
as it comes from the extrapolation of the data obtained for samples with finite quantum well width, in which the magnitudes of spin splitting are relatively large.

In addition to explaining the magnitude of spin splitting, the large  value of the $s$-$p$ exchange energy $J_{\mathrm{eh}}$ implied by our theory elucidates, as demonstrated recently,\cite{Astakhov:2008} why the spin relaxation time in GaAs:Mn can be by two orders of magnitude longer than that in GaAs containing a similar concentration of Ge acceptors.

When the bound hole concentration is diminished by donor compensation, the relative
importance of the $s$-$p$ exchange decreases. This can be the case of a ${\rm Ga}_{1-x} {\rm Mn}_x {\rm As}$ sample
with $x = 0.1 \%$, where  $N_0\alpha^{\mathrm{(app)}}   = +23$~meV, according to spin-flip Raman scattering.\cite{Heimbrodt:2001}
Even a lower value
$\left|N_0\alpha\right| = 14 \pm 4$~meV was found by  analyzing  the effect of the electrons on the Mn longitudinal relaxation time $T_1$ in
$n$-Ga$_{1-x}$Mn$_x$N with $ x \le 0.2$\%.\cite{Wolos:2003} The interpretation
of the data was carried out\cite{Wolos:2003} neglecting possible effects of the relaxation-time
bottleneck,\cite{Barnes:1981} which increases the apparent $T_1$.
It can be shown, however, that for the expected magnitudes of electron
spin-flip scattering times in wurtzite GaN:Mn,\cite{Sawicki:1986,Majewski:2005} this effect
leads to an underestimation of the $\left|N_0\alpha\right|$ by less than a factor of two.
On the other hand, as demonstrated in Sec.~VI,
the presence of positively charged donors shifts the electron wave function
away from negatively charged Mn acceptors, which results in a rather
strong reduction in the magnitude
of the apparent $s$-$d$ exchange integral in the relevant range of Mn
concentrations in $n$-(Ga,Mn)N and compensated (Ga,Mn)As.

\section{Model of the manganese acceptor}

The components $F_{\nu\mu}$ of the envelope function of the bound hole in the state
$\left| \mu \right>$, $\mu = \frac32, \frac12, -\frac12, -\frac32$,
are\cite{Bhattacharjee:2000}
\begin{eqnarray}
  \lefteqn{F_{\nu\mu}(\bm{r}) = \delta_{\mu\nu} R_0(r) Y_{00}(\theta, \phi) + {}} & & \\
    & & {} + \left< \frac32, \nu; 2, (\mu-\nu) \middle| \frac32, \mu \right> R_2(r) Y_{2,\mu-\nu}(\theta, \phi), \nonumber
\end{eqnarray}
where $\nu = \frac32, \frac12, -\frac12, -\frac32$ is the subband index ($j_z$).
Accordingly,
the spin-$\frac32$ angular-momentum matrices $j_\alpha$ act on the index $\nu$, while
$J_\alpha$ act on the index $\mu$.
The radial functions $R_0(r)$ and $R_2(r)$ are obtained from the Baldareschi-Lipari
equations by using a numerical solver of ordinary differential equations employing the standard values of the Luttinger parameters, $\gamma_1 = 6.85$, $\gamma_2 = 2.1$, $\gamma_3 = 2.9$, and $\kappa = 1.2$ (Ref.~\onlinecite{Dietl:2001}). In order to model the Mn acceptor in GaAs we take $\varepsilon_{\infty} = 10.66$ as the dielectric constant and the Gaussian central-cell potential with $r_0 = 2.8 \rm\,\hbox{\AA}$. The depth of the central-cell correction $V_0$ is chosen so that
the binding energy without the exchange contribution is $86.15 \rm\, meV$ (Ref.~\onlinecite{Bhattacharjee:2000}).
We will use
$f(r) = R_0(r)/\sqrt{4 \pi}$, $g(r) = R_2(r)/\sqrt{4 \pi}$, normalized as
$\int_0^\infty 4 \pi r^2 [f(r)^2+g(r)^2] \, dr = 1$. The functions $f(r)$, $g(r)$
are shown in Fig.~2.

Since both heavy and light hole masses are relevant,
the spatial decay of the bound hole wave function is not
characterized by a single exponent. An effective Mn acceptor
Bohr radius calculated from the participation ratio
is $0.76 \rm \, nm$ for the wave function determined above. This
agrees with the spatial extend of the probability density
observed by scanning tunneling microscopy for the hole
bound to Mn acceptor in GaAs.\cite{Yakunin:2004}

\begin{figure}[bt]
  \centering
  \includegraphics[width=0.95\columnwidth]{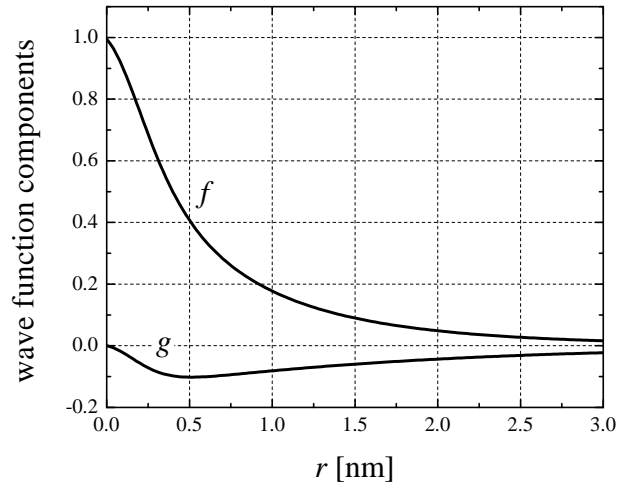}
  \caption{The components of the acceptor wave function (see the main body of the text for the definition of the functions $f$ and~$g$).}
  \label{fig2_eh}
\end{figure}

Now we calculate the Land\'e factor of the hole bound by the Mn acceptor, starting from the definition
of the magnetic moment,
$M_\alpha = - \bigl.\partial \mathcal{H}_{sph}/\partial B_\alpha \bigr|_{\bm{B} = 0}$,
where $\mathcal{H}_{sph}$ is the hole Hamiltonian in the spherical approximation,
\begin{eqnarray}
  \lefteqn{\mathcal{H}_{sph} = \frac{\hbar^2}{m} \Bigl\{ \frac12 \gamma_1 k^2
    - \bar\gamma \bigl[ (j_x^2 - \frac13 j^2) k_x^2 + c.p. \bigr] \Bigr.} & & \\
    & & \Bigl. {} - 2 \bar\gamma \bigl[ \{j_x, j_y \} \{ k_x, k_y \} + c.p. \bigr] \Bigr\}
    - \frac{e \hbar}{m} \kappa \bm{j} \cdot \bm{B}, \nonumber
\end{eqnarray}
in which $\{ A, B \} = \frac12 (AB + BA)$, $\bar\gamma = (2 \gamma_2 + 3 \gamma_3)/5$,
$k_\alpha = - i \frac{\partial}{\partial x_\alpha} - \frac{e A_\alpha}{\hbar}$,
and the vector potential in the axial gauge is
$A_\alpha = \varepsilon_{\alpha\beta\gamma} B_{\beta} x_{\gamma}/2$.
We have
\begin{eqnarray}
    \lefteqn{M_\alpha =
      \frac{e \hbar}{2 m} \Bigl\{ \gamma_1 \varepsilon_{\alpha\beta\gamma} x_\beta k_\gamma \Bigr.} & & \\
    & & {} - 2 \bar\gamma \bigl[ \{ j_\beta, j_\gamma \} - \frac13 \delta_{\beta\gamma} j^2 \bigr]
        \varepsilon_{\alpha\delta\gamma} x_\delta k_\beta \Bigr\}
      + \frac{e \hbar}{m} \kappa j_\alpha . \nonumber
\end{eqnarray}
We substitute $x_\alpha$ and $k_\alpha$ in the spherical coordinates:
$x = r \sin\theta \cos\phi$, $y = r \sin\theta \sin\phi$, $z = r \cos\theta$,
$k_x = - i (\sin\theta \cos\phi \, \partial_r
+ \frac1r \cos\theta \cos\phi \, \partial_\theta
- \frac{1}{r \sin\theta} \sin\phi \, \partial_\phi)$,
$k_y = - i (\sin\theta \sin\phi \, \partial_r
+ \frac1r \cos\theta \sin\phi \, \partial_\theta
+ \frac{1}{r \sin\theta} \cos\phi \, \partial_\phi)$,
$k_z = - i (\cos\theta \, \partial_r - \frac1r \sin\theta \, \partial_\theta)$.
Finally, by acting with $M_\alpha$ on $F_{\nu\mu}$
and performing the integration over $\theta$ and $\phi$ we obtain
\begin{eqnarray}
  \lefteqn{\left< \mu' \middle| M_\alpha \middle| \mu \right> =
    \frac{e \hbar}{2m} \frac45 \int_0^\infty 4 \pi r^2 \, dr \times {}} & & \\
  & & {} \times \Bigl\{ g(r) \bigl[(\gamma_1 - 2 \bar \gamma) g(r) - \bar \gamma r f'(r) \bigr] + {}
  \Bigr. \nonumber \\ & & \Bigl.
    {} + \bar\gamma f(r) \bigl[ 3 g(r) + r g'(r) \bigr]
  \Bigr\} J_{\alpha; \mu'\mu}
  + \frac{e \hbar}{m} \kappa \left< \mu' \middle| j_\alpha \middle| \mu \right>, \nonumber
\end{eqnarray}
where the value of the integral over $r$ is $-3.28$. Therefore,
$g_h = - [\frac45 \cdot (-3.28) + 2 \kappa \cdot 0.78] = 0.75$,
in a good agreement with the values given in Ref.~\onlinecite{Malyshev:1998}
(observe the opposite sign convention).
Moreover, substituting $g_1' = g_h$ and $g_2' = -0.07$ into the equation (3)
of Ref.~\onlinecite{Schneider:1987} yields $g_J = 2.80$, in a good agreement with
the experimental value of the complex $g$-factor, $g_J = 2.77$.

\section{Short-range $s$-$p$ exchange}

We now derive a form of the short-range $s$-$p$ exchange  interaction between a conduction-band electron and a hole, which is valid for any localization radius of the hole. We make use of the known value of the free exciton exchange splitting $\Delta$ and the exciton Bohr radius $a_X$.
Neglecting cubic terms of the form $j_x^3 s_x + j_y^3 s_y + j_z^3 s_z$,
we obtain for the Hamiltonian of the short-range interaction
a formula similar to Eq.~1 of Ref.~\onlinecite{Fu:1999},
\begin{equation}
  \mathcal{H}^x = - \frac12 \pi a_X^3 \Delta \, (\bm{s}\cdot\bm{j}) \, \delta(\bm{r}_h-\bm{r}_e),
\end{equation}
where $\Delta = 0.006 \, \rm meV$ (Ref.~\onlinecite{Blackwood:1994},
sign convention according to Ref.~\onlinecite{Ekardt:1977})
and $a_X = 12 \, \rm nm$.

In the present case, we consider the coupling of a conduction-band electron to a hole bound to the Mn acceptor. Hence, we calculate the matrix elements $\left< \bm{k}_e, \sigma'; \mu' \middle|
    \mathcal{H}^{x} \middle| \bm{k}_e, \sigma; \mu \right>$, where
$\bm{k}_e$ is the electron wave vector,
\begin{equation}
  \left< \bm{r}_e \middle| \bm{k}_e \right> = \frac{1}{\sqrt{\mathcal{V}}} e^{i \bm{k}_e \cdot \bm{r}_e},
\end{equation}
$\sigma$ is the electron spin, and $\mu$ numbers the spin states of the bound hole. Since
\begin{equation}
  \left< \bm{k}_e \middle| \delta(\bm{r}_h-\bm{r}_e) \middle| \bm{k}_e \right> =
    \frac{1}{\mathcal{V}}
\end{equation}
and
\begin{equation}
  \left< \mu' \middle| j_\alpha \middle| \mu \right> =
    \left( \left< f^2 \right> + \frac15 \left< g^2 \right> \right) J_{\alpha; \mu' \mu},
\end{equation}
where $\left< f^2 \right> = \int_0^\infty 4 \pi r^2 \, f(r)^2 \, dr$ etc.,
we obtain
\begin{equation}
  \mathcal{H}_{SR} = - \frac12 \Delta \frac{\pi a_X^3}{\mathcal{V}} \left( \left< f^2 \right> + \frac15 \left< g^2 \right> \right) \, \bm{s} \cdot \bm{J},
\end{equation}
casting the short-range interaction into the required form involving $\bm{J}$, not $\bm{j}$.
The numerical value is $\left< f^2 \right> + \frac15 \left< g^2 \right> = 0.78$.
A similar reduction factor of the acceptor splitting was obtained previously for a variational wave function.\cite{Mycielski:1983}
It appeared also in the case of DMS nanocrystals.\cite{Bhattacharjee:1995}
This value yields $-0.28 \rm\,eV$ as the contribution of the short-range interaction 
to $N_0 J_{\mathrm{eh}}$.

\section{Long-range $s$-$p$ exchange}

The long-range interaction operator is given by\cite{Pikus:1971}
\begin{eqnarray}
  \lefteqn{\mathcal{H}^{a}_{m'n',mn}(\bm{r}_1' \bm{r}_2', \bm{r}_1 \bm{r}_2) = {}} & & \\
    & & - \sum_{\alpha, \beta} Q^{\alpha\beta}_{m'\mathcal{K}n,\mathcal{K}n'm}
          \frac{\partial^2 V(\bm{r}_1 - \bm{r}_2')}{\partial r_{1\alpha} \partial r_{1\beta}}
          \, \delta(\bm{r}_1 - \bm{r}_2) \delta(\bm{r}_1' - \bm{r}_2'), \nonumber \\
  \lefteqn{Q^{\alpha\beta}_{m'\mathcal{K}n,\mathcal{K}n'm} =
    \frac{\hbar^2}{m^2 E_g^2} p^{\alpha}_{m'\mathcal{K}n'} p^{\beta}_{\mathcal{K}nm},} & &
\end{eqnarray}
where $V(\bm{r}) = e^2/(4 \pi \epsilon \epsilon_0 r)$ is the Coulomb potential.
In particular, for an exciton with momentum $\bm{K}$ in the spin state $j$, the matrix element
of this operator is
\begin{equation}
  \left< j' \bm{K}' \middle| \mathcal{H}^a \middle| j \bm{K} \right> =
    \frac{e^2}{\epsilon \epsilon_0} \frac{\hbar^2}{m^2 E_g^2}
    \phi_{j'} \phi_{j}^{*} \, \delta_{\bm{K}\bm{K}'},
\end{equation}
where $\phi_{j} = \sum_{mn} [\sum_{\alpha} n_{\alpha} p^{\alpha}_{m\mathcal{K}n}]^{*} f^{j}_{\bm{K}mn}(0)$,
$\bm{n} = \bm{K}/K$, and $f^{j}_{\bm{K}}(\bm{r})$ is the envelope function describing the relative motion
of the electron and the hole in an exciton in the state $\left| j \bm{K} \right>$.

We consider an exciton with the electron in the conduction band and the hole in the uppermost
valence band of the $\Gamma_8$ symmetry. The canonical basis for the latter is
$u_1 = \left| \frac32, \frac32 \right>$, $u_2 = \left| \frac32, \frac12 \right>$,
$u_3 = \left| \frac32, -\frac12 \right>$, $u_4 = \left| \frac32, -\frac32 \right>$,
\begin{eqnarray}
  u_1 & = & -\frac{1}{\sqrt{2}} (X + i Y) \uparrow, \\
  u_2 & = & \frac{1}{\sqrt{6}} [ -(X + i Y) \downarrow + 2 Z \uparrow ], \\
  u_3 & = & \frac{1}{\sqrt{6}} [ (X - i Y) \uparrow + 2 Z \downarrow ], \\
  u_4 & = & \frac{1}{\sqrt{2}} (X - i Y) \downarrow.
\end{eqnarray}
The time inversion operator acts as follows: $\hat\mathcal{K} u_i = \sum_j \mathcal{K}_{ji} u_j$, where
the matrix $\mathcal{K}$ is
\begin{equation}
  \mathcal{K} = \pmatrix{ 0& 0& 0& i \cr 0& 0& -i& 0 \cr 0& i& 0& 0 \cr -i& 0& 0& 0}.
\end{equation}
It is convenient to use instead of $\alpha = x, y, z$ the index $a = -1, 0, +1$, with
$n_{+1} = - (n_x + i n_y)/\sqrt{2}$, $n_0 = n_z$, $n_{-1} = (n_x - i n_y)/\sqrt{2}$.
Then we can express the momentum matrix elements in terms of Clebsch-Gordan coefficients:
\begin{eqnarray}
  \lefteqn{\sum_\alpha n_{\alpha} p^{\alpha}_{m\mathcal{K}n} = {}} & & \\
    & & P \sum_{n'} \sum_a \sqrt{\frac{4 \pi}{3}} Y_{1a}(\bm{n})
      \left< 1, a; \frac12, m \middle| \frac32, n' \right> \mathcal{K}_{n'n}, \nonumber
\end{eqnarray}
where $P = \left<S | P_z | Z\right>$. Assuming ground-state hydrogen wave-functions
for the envelope functions of the relative motion, we have
$f^{j}_{\bm{K}mn}(0) = (\pi a_X^3)^{-\frac12} \left< j \middle| \frac12, m; \frac32, n \right>$.
Again, Clebsch-Gordan coefficients have been used and $\left|j\right> = \left|J,J_z\right>$,
where $\bm{J}$ is the exciton spin ($J = 1, 2$). We have
$\phi_{\left| 2, J_z \right>} = 0$,
$\phi_{\left| 1, a \right>} = (2i/\sqrt{3}) \, P \, (\pi a_X^3)^{-\frac12} \, n_a$.
Therefore, $\| \bm{\phi} \|^2 = \frac43 \frac{P^2}{\pi a_X^3}$ and
we obtain the formula for the longitudinal-transverse exciton splitting,\cite{Ekardt:1977}
\begin{equation}
  \Delta_{LT} = \frac43 \frac{e^2}{\epsilon \epsilon_0} \frac{\hbar^2 P^2}{m^2 E_g^2} \frac{1}{\pi a_X^3}.
\end{equation}
Now we can express the strength of the long-range interaction in terms of $\Delta_{LT}$,
\begin{equation}
  Q^{\alpha\beta}_{m'\mathcal{K}n,\mathcal{K}n'm} =
    \frac34 \Delta_{LT} \pi a_X^3 \frac{\epsilon \epsilon_0}{e^2}
    \frac{p^{\alpha}_{m'\mathcal{K}n'} p^{\beta}_{\mathcal{K}nm}}{P^2},
\end{equation}
and calculate the matrix element
\begin{eqnarray}
  \lefteqn{\left< \bm{k}_e, \sigma'; \mu' \middle|
    \mathcal{H}^{a} \middle| \bm{k}_e, \sigma; \mu \right> = {}} & & \\
    & & \frac{1}{\mathcal{V}} \int d\bm{r}_1' \, d\bm{r}_2' \, d\bm{r}_1 \, d\bm{r}_2 \,
      e^{-i \bm{k}_e \bm{r}_1'} \, F_{\nu'\mu'}^{*}(\bm{r}_2') \times {} \nonumber \\
    & &  \quad {} \times \mathcal{H}^{a}_{\sigma'\nu',\sigma\nu}(\bm{r}_1' \bm{r}_2', \bm{r}_1 \bm{r}_2) \,
      e^{i \bm{k}_e \bm{r}_1} \, F_{\nu\mu}(\bm{r}_2) \nonumber.
\end{eqnarray}
\begin{figure}[bt]
  \centering
  \includegraphics[width=0.95\columnwidth]{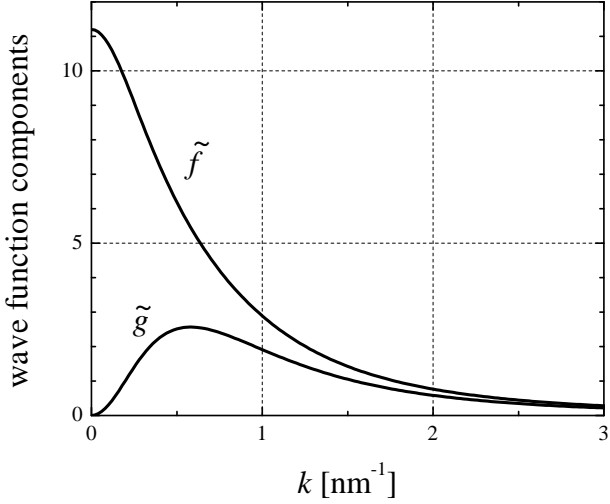}
  \caption{The components of the acceptor wave function in the momentum
  representation (see the main body of the text for the definition of the functions $\tilde f$ and $\tilde g$).}
  \label{fig3_eh}
\end{figure}
We use the following standard convention for the Fourier transform:
\begin{equation}
  \tilde f(\bm{k}) = \int d \bm{r} \, e^{-i \bm{k} \cdot \bm{r}} \, f(\bm{r}).
\end{equation}
Let
\begin{eqnarray}
  W_{\sigma' \nu, \nu' \sigma}(\bm{a}) & = & - Q^{\alpha\beta}_{\sigma' \nu, \nu' \sigma} \,
    \frac{\partial^2 V(\bm{a})}{\partial a_\alpha \partial a_\beta}, \\
  \tilde W_{\sigma' \nu, \nu' \sigma}(\bm{q}) & = &
    \frac34 \Delta_{LT} \pi a_X^3 \frac{q_{\alpha} q_{\beta}}{q^2}
    \frac{p^{\alpha}_{\sigma'\nu'} (p^{\beta}_{\sigma\nu})^{*}}{P^2}.
\end{eqnarray}
Then, using the properties of the Fourier transform, we can write the required matrix element as
\begin{eqnarray}
  \lefteqn{\left< \bm{k}_e, \sigma'; \mu' \middle|
    \mathcal{H}^{a} \middle| \bm{k}_e, \sigma; \mu \right> = {}} & & \\
    & & \quad \frac{1}{\mathcal{V}} \left\{ \tilde W_{\sigma' \nu, \nu' \sigma}
      \ast \left[(\tilde F_{\nu'\mu'})^{*} \cdot \tilde F_{\nu\mu}\right] \right\}(-\bm{k}_e), \nonumber
\end{eqnarray}
where $\ast$ denotes the convolution
\begin{equation}
  (\tilde f \ast \tilde g)(\bm{k}) =
    \int \frac{d \bm{q}}{(2 \pi)^3} \, \tilde f(\bm{q}) \, \tilde g(\bm{k} - \bm{q}).
\end{equation}
In particular, for $\bm{k}_e = 0$,
\begin{eqnarray}
  \lefteqn{\left< \sigma' \mu' \middle| \mathcal{H}^{a} \middle| \sigma \mu \right> =
    \frac34 \Delta_{LT} \frac{\pi a_X^3}{\mathcal{V}} \sum \int \frac{d \bm{q}}{(2 \pi)^3} \times {}} & & \\
    & & {} \times [\tilde F_{\nu'\mu'}(\bm{q})]^{*}
      \frac{q^a}{q} \left< 1, a; \frac12, \sigma' \middle| \frac32, \xi' \right> \mathcal{K}_{\xi', \nu'}
      \times {} \nonumber \\ & & {} \times
      \mathcal{K}_{\xi, \nu}^{*} \left< 1, b; \frac12, \sigma \middle| \frac32, \xi \right>
      \frac{(q^b)^{*}}{q} \tilde F_{\nu\mu}(\bm{q}), \nonumber
\end{eqnarray}
where the sum is over $a, b, \nu, \xi, \nu', \xi'$.
To calculate this integral, Fourier transforms of the envelope functions in the spherical coordinates are needed.
For $f(\bm{r}) = f(r) \, Y_{lm}(\bm{r}/r)$,
$\tilde f(\bm{k}) = \tilde f(k) \, Y_{lm}(\bm{k}/k)$, where
$\tilde f(k) = \int_0^\infty 4 \pi r^2 \, dr \, f(r) \, Q_l(k r)$, and
\begin{equation}
  Q_l(k r) = \frac12 \int_{-1}^{1} dx \, e^{-i k r x} P_l(x) = (-i)^l j_l(kr).
\end{equation}
For $l = 0$ and $2$,
\begin{eqnarray}
  Q_0(k r) & = & \frac{\sin k r}{k r}, \\
  Q_2(k r) & = & \frac{3 k r \cos k r + (k^2 r^2 - 3) \sin k r}{k^3 r^3}.
\end{eqnarray}
The functions $\tilde f(k)$ and $\tilde g(k)$ are shown in Fig.~3.
Now we can substitute the Fourier transforms into the integral
and separate the radial and the angular integration. 

By using the above results we obtain, 
\begin{equation}
  \mathcal{H}_{LR} =
    -\frac16 \Delta_{LT} \frac{\pi a_X^3}{\mathcal{V}}
      \left(\bigl< \tilde f^2 \bigr> - 2 \bigl< \tilde f \tilde g \bigr> + \bigl< \tilde g^2 \bigr> \right) \,
      \bm{s}\cdot\bm{J},
\end{equation}
where $\bigl< \tilde f^2 \bigr> = (2 \pi)^{-3} \int_0^\infty 4 \pi q^2 \, \tilde f(q)^2 \, dq$ etc.
To compute $\bigl< \tilde f \tilde g \bigr>$, one can use the identity ($r_1, r_2 > 0$)
\begin{eqnarray}
  \lefteqn{\frac{1}{(2 \pi)^3} \int_{k=0}^{\infty} 4 \pi k^2 \, dk \, Q_0(k r_1) Q_2(k r_2) = {} } \\
    & & \qquad \qquad \nonumber
    \frac{1}{4 \pi} \left[ \frac{\delta(r_2-r_1)}{r_1 r_2} - \frac{3 \theta(r_2-r_1)}{r_2^3} \right],
\end{eqnarray}
from which it follows immediately that
\begin{equation}
  \bigl< \tilde f \tilde g \bigr> = \left< f g \right>
    - 12 \pi \int_{0}^{\infty} dr_2 \int_{0}^{r_2} dr_1 \, \frac{r_1^2}{r_2} f(r_1) g(r_2).
\end{equation}
The numerical values are $\Delta_{LT} = 0.08 \pm 0.02 \, \rm meV$ (Refs.~\onlinecite{Ekardt:1979} and~\onlinecite{Ulbrich:1977}) and
\begin{equation}
  \frac16
  \left(\bigl< \tilde f^2 \bigr> - 2 \bigl< \tilde f \tilde g \bigr> + \bigl< \tilde g^2 \bigr> \right) = 0.024.
\end{equation}
Hence, the contribution of the long-range $s$-$p$ interaction to $N_0 J_{\mathrm{eh}}$ is $-0.23 \rm\,eV$, of the same order
as the short-range part. Taking into account experimental uncertainty of the relevant parameters
(we assume a $1 \rm\, nm$ error of $a_X$ and a $2 \rm\, \mu eV$ error of $\Delta$) we
obtain the total magnitude of the electron-hole exchange energy  $N_0 J_{\mathrm{eh}} = -0.51 \pm 0.17 \, \rm eV$.

For $\bm{k}_e \ne 0$, the spherical symmetry is broken and $\mathcal{H}_{LR}$ can no longer be cast
into the form $\bm{s} \cdot \bm{J}$.

\section{Apparent $s$-$d$ exchange in $n$-type case}

So far we have considered $p$-type systems, in which
the interaction of photoelectrons with neutral Mn complexes is relevant. Now we examine compensated III--V Mn-based DMSs, in which the electron concentration $n$ is greater than that of Mn impurities. In such samples of GaN:Mn and GaAs:Mn, strongly reduced magnitudes of the $s$-$d$ exchange integral have been found  by electron spin resonance,\cite{Wolos:2003} and by spin-flip Raman scattering,\cite{Heimbrodt:2001} respectively.

We note that in such samples  Mn acceptors are ionized. Also ionized are donors, as the electron concentration corresponding to an insulator-to-metal transition is relatively low in the case of the conduction-band carriers. The presence of the corresponding repulsive and attractive Coulomb interactions means that
the probability of finding a conduction-band electron at the core
of the magnetic ion is reduced, and hence the \emph{apparent} value of the
exchange energy (the observed spin splitting) is diminished. It is worth
noting that the possibility that the  Coulomb potentials could affect the apparent
value of the exchange integrals has already been mentioned in the context
of divalent Mn in GaN,\cite{Wolos:2003} and trivalent Fe in HgSe.\cite{Wilamowski:1988}

To evaluate a lower limit of the effect we neglect the
presence of compensating donors and calculate the apparent $s$-$d$ exchange
integral $\alpha^{\mathrm{(app)}}$ for an electron subject to the repulsive potential generated
by the unoccupied Mn acceptors. We follow a Wigner-Seitz-type approach put forward by
Benoit \`a la Guillaume \emph{et al.}\cite{Benoit:1992} 
to describe the interaction of the carrier spin with
the Mn ions in the case of the strong-coupling limit, that is when the
depth of the local Mn potential is comparable to the carrier bandwidth. It has
been found in the subsequent works\cite{Tworzydlo:1994,Tworzydlo:1995} that the corrections to the
Wigner-Seitz approach caused by a random distribution of Mn ions are
quantitatively unimportant.

We consider a  Mn ion with the $5/2$ spin $\vec S_i$
located at $\vec R_i$, which interacts with
the carrier via the Heisenberg term $I(\vec r
- \vec R_i)\vec s \cdot \vec S_i $.
The form of the function $I(\vec r - \vec R_i)$ makes
the interaction local: it vanishes outside the core of the Mn
ion. For simplicity, $I(\vec r - \vec R_i) = a \, \theta(b - |\vec r -
\vec R_i|)$. The exchange energy is then $\alpha = \int d^{3} \vec r
\, I(\vec r) = a \cdot \frac{4}{3} \pi b^3$. Moreover, in case of
III--V compounds considered here, the impurity generates an electrostatic
potential. If screening by the electrons is present, as in case of $n$-${\rm
  Ga}_{1-x}{\rm Mn}_x{\rm N}$, this potential is
$e^2\exp(-\lambda r)/(4\pi\varepsilon \varepsilon_0 r)$,
where $\varepsilon$ is the static dielectric constant,
and the screening parameter $\lambda$ is given by
$\lambda^2 = e^2 \mathcal{N}(\mathcal{E}_F)/(\varepsilon_0 \varepsilon )
$, where $\mathcal{N}(\mathcal{E}_F) =
\frac{3}{2} n/k T_F$ (see Ref.~\onlinecite{Ziman:1972}, \S 5.2).
For the ${\rm Ga}_{1-x}{\rm Mn}_x{\rm N}$ samples,\cite{Wolos:2003} $n
\approx 10^{19}\,{\rm cm}^{-3}$ corresponds to $T_F \approx 890 \, \rm
K$ ($\mathcal{E}_F \approx 0.12 \, \rm eV$), and therefore $1/\lambda
\approx 1.6 \, \rm nm $.

\begin{figure}[tb]
\includegraphics[width=0.98\columnwidth]{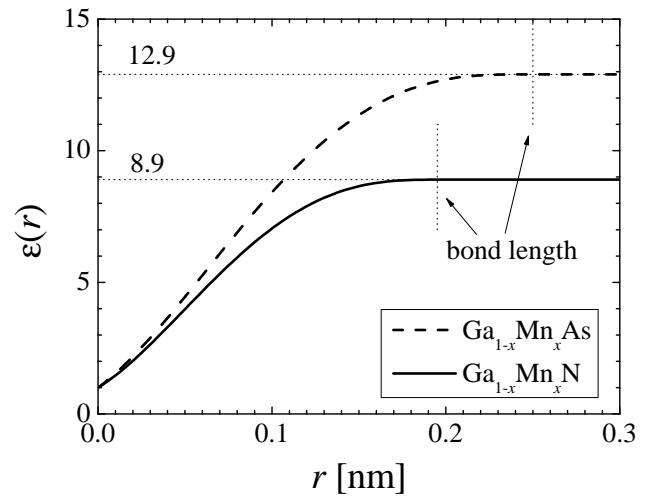}%
\caption[]{The assumed dependence of the dielectric constant $\varepsilon$
on the distance $r$ to an ionized acceptor.}
\label{epsiloncurve}
\end{figure}

In the spirit of the Wigner-Seitz approach we assume that the carrier
energy~$E$ and the envelope function~$\psi(r)$ are given by the ground
state $s$ solution of the one-band effective-mass equation which
contains the potential $U(r)$ created by the magnetic ion located at
$r = 0$. The standard one-impurity boundary condition $\psi(r) \to 0$
for $r \to \infty$ is replaced by the matching condition $\psi'(r) =
0$ at $r = R$ to take into account the presence of other magnetic
ions. The value~$R$ is determined by the concentration of the magnetic
ions~$x$ according to the equation $(4 \pi R^3/3)^{-1} = N_0 x$.
The exchange interaction is modeled by a square-well
potential $U \theta(b-r)$ superimposed on the electrostatic potential
of an elementary charge located at $r = 0$. The potential $U = \pm \frac54 a$ is, of
course, different for spin-down and spin-up carriers.

We first ignore free-carrier screening, $\lambda \to 0$. The solution of
the time-independent Schr\"odinger equation for the conduction band
electron in then
\begin{equation}
  \label{eqpsi1}
  \psi(r) = c_0 \exp(-\beta r) \Phi(1+\frac{A}{\beta}; 2; 2 \beta r) \equiv c_0 f
\end{equation}
for $0 < r < b$, and the following linear combination for $b < r < R$
\begin{eqnarray}
  \psi(r) & = & c_1 \exp(-\beta' r) \Psi(1+\frac{A}{\beta'}; 2; 2 \beta' r) + {} \nonumber
    \\
    & & {} + c_2 \exp(\beta' r) \Psi(1-\frac{A}{\beta'}; 2; -2 \beta'
    r) \nonumber \\ & \equiv & c_1 g + c_2 h, \label{eqpsi2}
\end{eqnarray}
where $A = e^2 m^{*}/(4\pi \varepsilon \varepsilon_0 \hbar^2)$,
$\beta = [2 m^{*}(U-E)]^{\frac12}/\hbar$, $\beta' = [2
m^{*}(-E)]^{\frac12}/\hbar$ (notice that changing the sign of $\beta$
leaves $\psi$ invariant, while changing the sign of $\beta'$
interchanges $c_1$ with $c_2$; also, $\Phi$ and $\Psi$ are not in
general linearly independent). We used the symbols $\Phi$, $\Psi$ for
the confluent hypergeometric functions ${}_{1}F_{1}(a; b; z)$, $U(a;
b; z)$ (Ref.~\onlinecite{Slater:1960}). The constants $c_0$, $c_1$, $c_2$ are determined
by the continuity conditions $\psi(b^{-}) = \psi(b^{+})$,
$\psi'(b^{-}) = \psi'(b^{+})$. Solving those two equations we obtain
an equation for~$E$,
\begin{equation}
  \label{eqdpsiReq0}
  \frac{w_{f, h}(b) g'(R) - w_{f, g}(b) h'(R)}{w_{g, h}(b)} = 0,
\end{equation}
where by $w_{f, g}$ we denoted the Wronskian $f g' - f' g$.

We assume the following parameters for ${\rm Ga}_{1-x}{\rm Mn}_x {\rm
  N}$: $m^{*} = 0.22 \, m_e$, $N_0 = 4.38 \times 10^{22} \, {\rm cm}^{-3}
= 0.006495 \, \rm a.u.$, $\varepsilon = 8.9$; and the following for
${\rm Ga}_{1-x}{\rm Mn}_x {\rm As}$: $m^{*} = 0.067 \, m_e$, $N_0 =
2.21 \times 10^{22} \, {\rm cm}^{-3} = 0.003281 \, \rm a.u.$,
$\varepsilon = 12.9$. In the experiments, samples were used with $0.01\% \le x \le 0.2\%$ of
$\rm Mn$ in $\rm GaN$,\cite{Wolos:2003} and with $0.0006\% \le x \le 0.03\%$
of $\rm Mn$ in $\rm GaAs$.\cite{Myers:2005} Those concentrations
correspond to $R$ up to about $75 \, \rm a.u.$ for $\rm GaN$ and up to
about $250 \, \rm a.u.$ for $\rm GaAs$.

To visualize the effect of the Coulomb term in the
Mn potential, we have calculated the energies and wave functions
including the additional Coulomb term for both
GaN ($b = 2 \, {\rm a.u.} \approx 0.1 \, \rm nm$, $a = 0.0371 \, \rm
a.u. = 1.0 \rm\, eV$) and $\rm GaAs$ ($b = 2 \, \rm a.u. \approx 0.1
\, \rm nm$, $a = 0.0735 \, {\rm a.u.} = 2.0 \, \rm eV$).
These parameters correspond to $N_0 \alpha = 0.22 \,
\rm eV$. We have found that
when calculating $\alpha^{\mathrm{(app)}}/\alpha$, the
details of the exchange potential (like the values of~$b$ and $\alpha$
within the expected range) are not
quantitatively important.

\begin{figure}[tb]
\includegraphics[width=0.915\columnwidth]{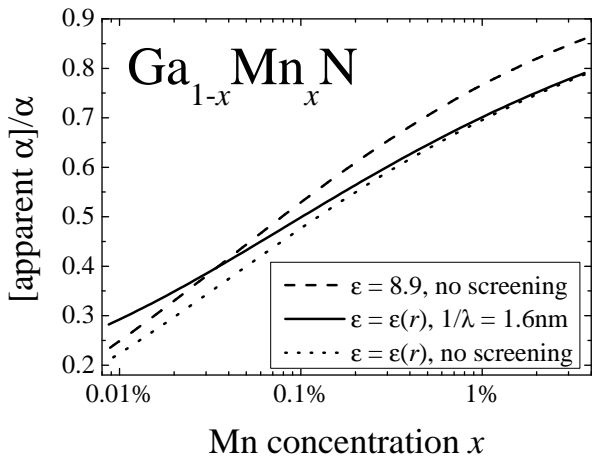}%
\caption[]{The dependence of the ratio of the apparent and bare exchange energies
$\alpha$ on~$x$ for for ${\rm Ga}_{1-x}{\rm Mn}_x {\rm N}$ and various models of screening.}
\label{xdepN}
\end{figure}

\begin{figure}[tb]
\includegraphics[width=0.98\columnwidth]{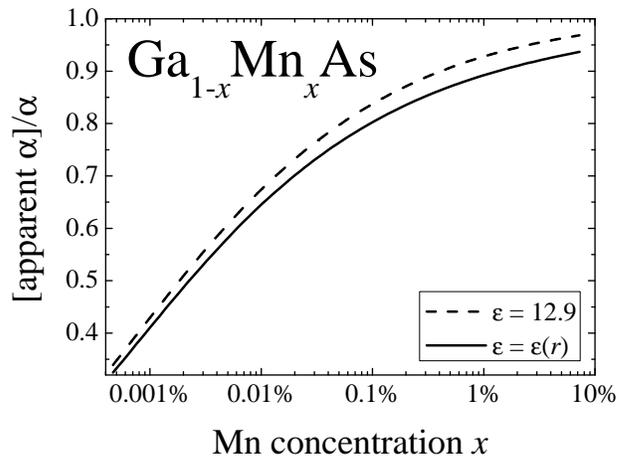}%
\caption[]{The dependence of the ratio of the apparent and bare exchange energies
$\alpha$ on~$x$ for ${\rm Ga}_{1-x}{\rm Mn}_x {\rm As}$.}
\label{xdepAs}
\end{figure}

In order to take into account the fact that the core and lattice
polarizability decrease at small distances, $\varepsilon \rightarrow 1$ for $r \rightarrow 0$,
we interpolate $\varepsilon(r)$ between $\varepsilon(0) = 1$ and the macroscopic value
attained at a distance of the bond length. The
assumed dependence, presented in Fig.~\ref{epsiloncurve}, is similar to
that of the Thomas-Fermi model.\cite{Resta:1977}
When  $\varepsilon = \varepsilon(r)$ and/or free-carrier screening is included,
we find the solution $\psi(r)$ of the Schr\"odinger equation for
the given potential $U(r)$ numerically, as the Eqs.~(\ref{eqpsi1})
and~(\ref{eqpsi2}) are only valid for the Coulomb
potential. Then, the spin splitting for a given value of~$x$ (or for
the corresponding~$R$) is evaluated as the difference of the
energy~$E$ calculated for the spin-up and spin-down carriers from the
equation $\psi'(R) = 0$. Here, $\psi(r)$ is the numerical solution of
the Schr\"odinger equation with the potential that is different for
spin-up and spin-down carriers.

The results of our calculations of $\alpha^{\mathrm{(app)}}/\alpha$ as a function
of the Mn ion concentration $x$ are presented in Fig.~\ref{xdepN}
(${\rm Ga}_{1-x}{\rm Mn}_x {\rm N}$) and in Fig.~\ref{xdepAs} (${\rm
Ga}_{1-x}{\rm Mn}_x {\rm As}$). Independently of assumptions
concerning screening, in both materials  $\alpha^{\mathrm{(app)}}/\alpha$ diminishes
significantly when $x$ decreases, up to factor of 3 in the experimentally relevant
range of~$x$. However, this reduction of $\alpha^{\mathrm{(app)}}/\alpha$ is still smaller
than that seen experimentally,\cite{Heimbrodt:2001,Wolos:2003} presumably because of an additional
effect coming from the presence of attractive potentials brought about by
compensating nonmagnetic donors.

\section{Summary and outlook}

In order to understand the magnitude of the spin splitting of photoelectrons in Mn-based III--V DMSs, we have developed theory of the $s$-$p$ exchange interaction between conduction-band electrons
and holes localized on Mn acceptors, taking into account both short- and long-range contributions. According to our results, this exchange overcompensates the
$s$-$d$ interaction of the electrons with the Mn spins, making the resulting coupling to
be antiferromagnetic. The theory describes, employing the standard value of the $s$-$d$ exchange energy $N_0\alpha = 0.22$~eV, the recent results on spin splitting\cite{Myers:2005,Poggio:2005,Stern:2007} and spin relaxation time\cite{Astakhov:2008}
of photoelectrons
in GaAs:Mn with low Mn concentrations.

In view of our work, it would be remarkable
to carry out Zeeman spectroscopy on nonmagnetic $p$-type semiconductors on the insulating side
of the insulator-to-metal
transition, where a large exchange splitting of the conduction band by the bound holes is predicted by the present theory. It would also be interesting to put forward an {\em ab initio} approach capturing such an effect.  Finally, we note that the confinement-induced changes in the symmetry of the electron wave function explain,\cite{Dalpian:2006,Stern:2007} {\em via} the $sp$-$d$ kinetic exchange, the corresponding experimentally-revealed growth of the antiferromagnetic contribution to the exchange
integral.\cite{Myers:2005,Poggio:2005,Stern:2007} The question about the role of the simultaneously appearing $p$-$p$ exchange is opened to further studies.

Furthermore, we have considered the interaction of conduction-band electrons with Mn ions in compensated $n$-type III--V DMSs, taking into account the electrostatic potential created by the
magnetic ions. A substantial reduction in the magnitude
of the apparent exchange energy has been found at low Mn concentrations,
and interpreted as
coming from the decrease of the carrier probability density at the
core of the magnetic ion caused by the electrostatic repulsion. It has
been suggested that this effect, enhanced by an attractive potential of
compensating donors, accounts for reduced values of the
exchange spin splitting observed experimentally in compensated III--V DMSs containing
a minute amount of Mn.\cite{Heimbrodt:2001,Wolos:2003} In view
of our findings, the presence of electrostatic potentials associated with magnetic
ions makes that
the apparent exchange energies should not be viewed as universal but rather dependent on the
content of the magnetic constituent and compensating donors.

\begin{acknowledgments}
  We would like to thank W. Bardyszewski and R. Buczko for discussions.
  This work was supported in part by the EC project NANOSPIN (Grant No. FP6-2002-IST-015728).
\end{acknowledgments}


\end{document}